\documentclass[sigconf]{acmart}
\usepackage{algorithmicx}
\usepackage{url}
\usepackage{bbm}
\usepackage{bm}
\usepackage{subcaption}
\usepackage[export]{adjustbox}
\usepackage{amsmath}
\usepackage{mathtools}
\usepackage{multirow}
\usepackage{makecell}
\usepackage{amsthm}
\usepackage{algpseudocode}
\usepackage{balance}
\usepackage{xcolor}
\usepackage[linesnumbered,ruled,vlined]{algorithm2e}

\SetCommentSty{mycommfont}

\SetKwInput{KwInput}{Input}                
\SetKwInput{KwOutput}{Output}           
\SetKwInput{KwInit}{Initialization}

\AtBeginDocument{%
  \providecommand\BibTeX{{%
    \normalfont B\kern-0.5em{\scshape i\kern-0.25em b}\kern-0.8em\TeX}}}

\copyrightyear{2024}
\acmYear{2024}
\setcopyright{acmlicensed}
\acmISBN{XXXXXXXXXXXXX}
\acmDOI{10.1145/XXXXXXXXXX}

\begin{document}
\title{Triple Modality Fusion: Aligning Visual, Textual, and Graph Data with Large Language Models for Multi-Behavior Recommendations}


\renewcommand{\shortauthors}{Luyi Ma et al.}

\author{Luyi Ma}
\authornote{All three authors contributed equally to this research.}
\affiliation{%
  \institution{Walmart Global Tech}
  \city{Sunnyvale}
  \state{California}
  \country{USA}}
\email{luyi.ma@walmart.com}

\author{Xiaohan Li}
\authornotemark[1]
\affiliation{%
  \institution{Walmart Global Tech}
  \city{Sunnyvale}
  \state{California}
  \country{USA}}
\email{xiaohan.li@walmart.com}

\author{Zezhong Fan}
\authornotemark[1]
\affiliation{%
  \institution{Walmart Global Tech}
  \city{Sunnyvale}
  \state{California}
  \country{USA}}
\email{zezhong.fan@walmart.com}

\author{Kai Zhao}
\affiliation{%
  \institution{Walmart Global Tech}
  \city{Sunnyvale}
  \state{California}
  \country{USA}}
\email{kai.zhao@walmart.com}

\author{Jianpeng Xu}
\affiliation{%
  \institution{Walmart Global Tech}
  \city{Sunnyvale}
  \state{California}
  \country{USA}}
\email{jianpeng.xu@walmart.com}

\author{Jason Cho}
\affiliation{%
  \institution{Walmart Global Tech}
  \city{Sunnyvale}
  \state{California}
  \country{USA}}
\email{jason.cho@walmart.com}

\author{Praveen Kanumala}
\affiliation{%
  \institution{Walmart Global Tech}
  \city{Sunnyvale}
  \state{California}
  \country{USA}}
\email{pkanumala@walmart.com}

\author{Kaushiki Nag}
\affiliation{%
  \institution{Walmart Global Tech}
  \city{Sunnyvale}
  \state{California}
  \country{USA}}
\email{kaushiki.nag@walmart.com}

\author{Evren Korpeoglu}
\affiliation{%
  \institution{Walmart Global Tech}
  \city{Sunnyvale}
  \state{California}
  \country{USA}}
\email{EKorpeoglu@walmart.com}

\author{Sushant Kumar}
\affiliation{%
  \institution{Walmart Global Tech}
  \city{Sunnyvale}
  \state{California}
  \country{USA}}
\email{sushant.kumar@walmart.com}

\author{Kannan Achan}
\affiliation{%
  \institution{Walmart Global Tech}
  \city{Sunnyvale}
  \state{California}
  \country{USA}}
\email{kannan.achan@walmart.com}
\begin{abstract}
    Integrating diverse data modalities is crucial for enhancing the performance of personalized recommendation systems. Traditional models, which often rely on singular data sources, lack the depth needed to accurately capture the multifaceted nature of item features and user behaviors. This paper introduces a novel framework for multi-behavior recommendations, leveraging the fusion of triple-modality, which is visual, textual, and graph data through alignment with large language models (LLMs). By incorporating visual information, we capture contextual and aesthetic item characteristics; textual data provides insights into user interests and item features in detail; and graph data elucidates relationships within the item-behavior heterogeneous graphs. Our proposed model called Triple Modality Fusion (TMF) utilizes the power of LLMs to align and integrate these three modalities, achieving a comprehensive representation of user behaviors. The LLM models the user's interactions including behaviors and item features in natural languages. Initially, the LLM is warmed up using only natural language-based prompts. We then devise the modality fusion module based on cross-attention and self-attention mechanisms to integrate different modalities from other models into the same embedding space and incorporate them into an LLM. Extensive experiments demonstrate the effectiveness of our approach in improving recommendation accuracy. Further ablation studies validate the effectiveness of our model design and benefits of the TMF. 
\end{abstract}



\maketitle

\section{Introduction}
In the rapid development of personalized recommendation systems, the ability to leverage diverse data modalities is pivotal for delivering accurate and relevant recommendations to users~\cite{he2016vbpr, zheng2017joint}.  Traditional recommendation models often rely on singular data sources, which limits their capability to fully comprehend the intricate and multifaceted nature of user behaviors and item features. This paper addresses this limitation by introducing a novel framework for multi-behavior recommendations that leverages the fusion of three distinct modalities: visual, textual and graph data based on the power of Large Language Models (LLMs).

Recently, inspired by the great success of Large Language Models (LLMs)~\cite{brown2020language, chowdhery2023palm} and Multi-Modal Language Models~\cite{team2023gemini}, exploring the potential of LLMs and VLMs in 
recommendation systems are attracting attention~\cite{liao2023llara, cui2022m6, bao2023tallrec, geng2022recommendation, hou2024large, liu2023chatgpt, tan2024idgenrec, geng2023vip5}, especially driven by extensive world knowledge and
reasoning capabilities of LLMs~\cite{brown2020language}. At the core is to reshape sequential
recommendation as the language modeling task — that is, convert
users' behavioral sequence into the textual input prompt. For example, as illustrated in Figure~\ref{tmf-framework}, "The user \textbf{views} \textit{grey athletic T-shirt}, \textbf{adds to cart} \textit{water bottle}, what's the next to \textbf{purchase}?" is the prompt we input to the LLM, where the bold words represent user behaviors and italic words are item names. In this example, we explicitly input the behaviors in the prompt to predict users’ future interactions on the target behavior (purchase in the example) with the help of auxiliary behaviors (view and add to cart).  As user behaviors indicate the user's level of interest in the item based on past studies~\cite{jin2020multi, xia2021graph, yang2022multi, xia2022multi}, to make use of all types of behaviors, we propose a multi-modality fusion model based on LLM for multi-behavior recommendations.

Existing multi-behavior recommendation systems (MBRS) primarily explore two approaches: 1) sequence-based MBRS~\cite{su2023personalized, yuan2022multi, xia2022multi}, which employ sequential models like the Transformer~\cite{vaswani2017attention} to capture user representations from their behavior patterns in interaction sequences;  2) graph-based MBRS~\cite{xia2021graph, jin2020multi, xuan2023knowledge, yang2022multi}, which represent behaviors as nodes or edges in a graph structure to reflect relationship strengths. Recent advancements in LLMs have shown their effectiveness in sequential recommendation systems.~\cite{cui2022m6, geng2022recommendation, harte2023leveraging, yang2024sequential}. However, there has been no integration of graph structures within MBRS using LLMs. In response, we leverage an LLM as the backbone recommendation model, incorporating graph modality embeddings into the LLM. Additionally, we align the LLM with visual and textual item modalities, as item images and textual descriptions play a crucial role in understanding item characteristics~\cite{he2016vbpr, yu2018aesthetic, li2023text}.

Our proposed model, named Triple Modality Fusion (TMF), aligns and integrates visual, textual and graph data into an LLM-based Recommender. Including visual information captures contextual and aesthetic item characteristics. Textual data provides detailed insights into user behaviors and item features. Furthermore, we model the user behaviors into an item-behavior graph, which offers a structural perspective that is crucial for modeling complex interactions. The TMF framework operates by initially warming up the LLM-based recommender using natural language-based prompts, ensuring that the model develops a strong foundational understanding of user behaviors and item features. We also introduce a modality fusion module based on cross-attention and self-attention mechanisms designed to project different modalities from other models into a unified embedding space, facilitating seamless integration with the LLM. This approach enables the TMF model to effectively align image, text, and graph data, resulting in robust and holistic representations of user behaviors and items. Extensive experiments conducted on three benchmark datasets demonstrate the superior performance of our TMF model in improving recommendation accuracy. Our model is also deployed in a real-world recommendation production to generate candidate set to ranking models in an offline manner. Based on human evaluation, we find the generated results from the TMF framework align better compared to the state-of-the-art model.

In summary, our contributions in this paper are listed below:
\begin{itemize}
    \item We introduce the Triple Modality Fusion framework, which aligns visual, textual, and graph data through integration with an LLM-based recommender.
    \item We propose a modality fusion module based on cross-attention and self-attention mechanisms to fuse all three modalities in the same embedding space and then can be seamlessly integrated into the LLM.
    \item The results validate the effectiveness of our model design and underscore the benefits of incorporating triple modality fusion in MBRS. The model is also used in real-world production.
\end{itemize}

\section{Related Works}
\subsection{Multi-Behavior Recommendation}
To enhance the capability of recommendation systems to capture user preferences more effectively, various approaches have been developed that incorporate multi-behavior information about users and items. Multi-behavior recommendation systems (MBRS), aiming to bolster the recommendation of a target behavior by leveraging auxiliary behavioral data, can be generally classified into two categories. 1) For sequence-based MBRS, they model behavior sequences as inputs into a sequential model. Yuan et al.~\cite{yuan2022multi} applies the transformer model~\cite{vaswani2017attention} to model heterogeneous item-level multi-behavior dependencies. The study in~\cite{xia2022multi} explores the interconnections among behavior sequences using a temporal graph transformer. 
2) For graph-based MBRS, they utilize Graph Neural Networks (GNNs)~\cite{li2020dynamic, liu2023group, li2022time} and Knowledge Graphs (KGs)~\cite{liu2020basket, li2021pre} to model behaviors. the MBGCN model~\cite{jin2020multi} applies a graph convolutional network to learn the similarities in behaviors and understand user interests. KMCLR~\cite{xuan2023knowledge} applied contrastive learning on a KG to fully use multi-behavior information. Furthermore, MB-GMN~\cite{xia2021graph} incorporates graph meta-learning into
the multi-behavior pattern modeling. MBHT~\cite{yang2022multi} utilizes hypergraph to learn both short-term and long-term cross-type behavior dependencies.  

Our proposed Triple Modality Fusion (TMF) improves the performance of sequence-based MBRS with the power of LLMs, which are effective in recommendation systems~\cite{cui2022m6, geng2022recommendation}. Traditional recommendation models often rely on singular data
modality~\cite{koren2021advances, li2022mitigating}. As graph modality has been demonstrated useful in MBRS, we incorporate the embeddings learned from a graph-based model to LLMs with an MLP adaptor. Moreover, as item images are also important features~\cite{he2016vbpr, deldjoo2020recommender, forouzandehmehr2023character}, we also align the visual representations of items with LLMs with the help of Vision Transformer~\cite{DBLP:conf/iclr/DosovitskiyB0WZ21}.

\subsection{LLM-based Recommendation Systems}
Recently, Large Language Models (LLMs) have demonstrated their advantages in diverse applications~\cite{chen2024relation, fang2024llm}. Research on LLM-based recommendation systems falls into three distinct categories: representation learning, in-context learning, and generative recommendation. In representation learning of LLM-based recommendation systems, the primary use of LLMs is to enhance user and item representations by utilizing contextual data during the recommendation process~\cite{hou2022towards, li2023text, qiu2021u, yao2022reprbert, zhang2022gbert}. 
In-context learning of LLM-based recommendation systems have
concentrated on probing the potential of LLMs such as ChatGPT~\footnote{https://openai.com/chatgpt/} to directly generate recommendations without the need for training or fine-tuning~\cite{gao2023chat, hou2024large, liu2023chatgpt, sun2023chatgpt, dai2023uncovering}. 
However, ChatGPT cannot generate recommendations with accuracy competitive with that of traditional
recommendation methods for all tasks based on the research from~\cite{liu2023chatgpt}.

This paper focuses on the last category: LLM-based generative recommendation. This novel approach transitions from the traditional ranking-based recommendation to a pure text-to-text method. So it can directly generate the target item in natural language. P5~\cite{geng2022recommendation} and its extension VIP5~\cite{geng2023vip5} introduced a generative recommendation model on five different tasks, leveraging a pre-trained T5 as the recommendation backbone. M6-Rec~\cite{cui2022m6} fine-tunes a large-scale industrial pre-trained LLM on recommendation tasks. LLaRA~\cite{liao2023llara} uses a hybrid prompting method that integrates
item embeddings learned by traditional recommendation models. IDGenRec~\cite{tan2024idgenrec} trains a textual ID generator integrated with the LLM-based recommender. In our proposed TMF framework, we further enhance recommendation performance by integrating the graph modality into the LLM-based recommendation model.

\begin{figure*}
    \centering
    \includegraphics[width=0.95\textwidth]{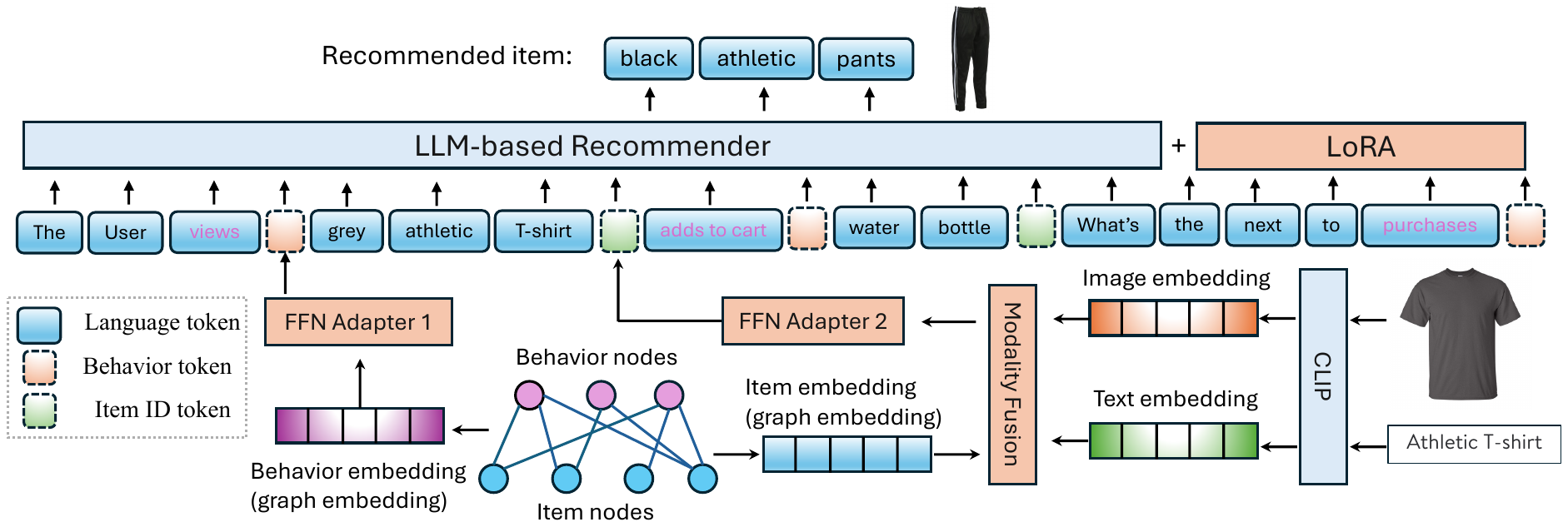}
    \caption{The Triple Modality Fusion (TMF) framework for multi-behavior recommendation. The blue squares are frozen models, and the red models are open to training in the training steps.}
    \label{tmf-framework}
\end{figure*}

\begin{figure}
    \centering
    \includegraphics[width=0.4\textwidth]{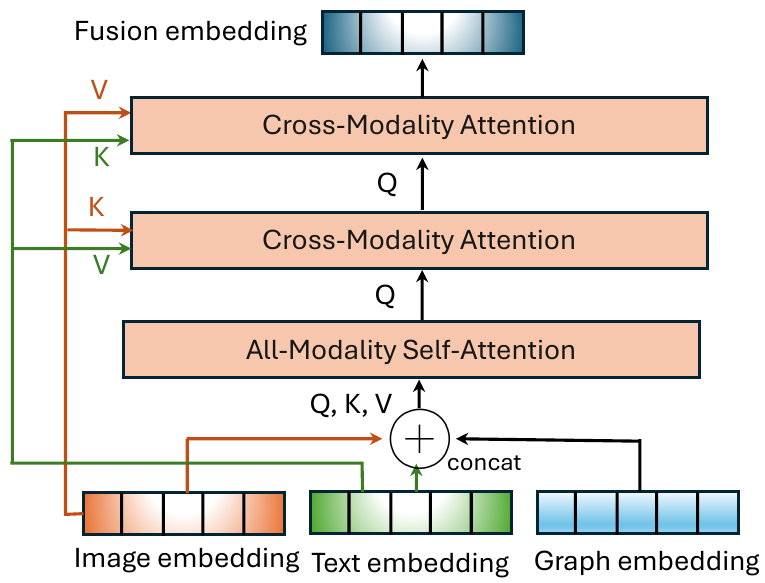}
    \caption{The Modality Fusion module in Figure~\ref{tmf-framework}.}
    \label{att}
\end{figure}

\section{Preliminary} \label{preliminary}
In this section, we introduce the preliminary information about our proposed TMF. 
\subsection{Problem Definition}
In multi-behavior recommendation systems (MBRS), we distinguish between target behaviors, the types of interactions we seek to predict, and auxiliary behaviors, which supplement the recommendation process. Auxiliary behaviors provide context by revealing different user preferences to enhance the performance of the recommendation on target behavior types. For instance, on many e-commerce platforms, purchasing behaviors are often targeted for prediction because of their direct correlation with the Gross Merchandise Volume (GMV), representing the total sales value of merchandise~\cite{fang2024llm, wu2018turning}. We define our approach to the multi-behavior recommendation problem as follows:
\begin{itemize}
    \item \textbf{Input:} The images and descriptions of items; The interaction history of a given user $u$ consisting of $n$ interacted items and their associated behaviors $H_{u} = {(i_1, b_1), (i_2, b_2), ..., (i_n, b_n)}$, where $i$ represent item and $b$ means behavior.
    \item \textbf{Output:} Text tokens representing the recommended item.
\end{itemize}

\subsection{Visual and Textual Modalities}
In modern recommendation systems, the item images and descriptions are important modalities for the users to understand the details of items~\cite{he2016vbpr, yu2018aesthetic, li2023text}. Given an item $i$, we define its visual embedding $v_{image}^{i}$ and textual embedding $v_{text}^{i}$ by Eq. \ref{image_embedding_of_item} and \ref{text_embedding_of_item} by the pre-trained image encoder $\text{F}_{visual}(\cdot)$ and the text encoder $\text{F}_{text}(\cdot)$ respectively, where $image_{i}$ is the image of the item $i$, and $text_{i}$ represents the textual information of the item $i$. Specifically, we concatenate textual information for the item $i$, including title, brand, categories, and description, to build $text_{i}$. In most cases, $text_{i}$ and $image_{i}$ are highly correlated to ensure the fidelity of item information. 
\begin{equation}
    v_{image}^{i} = \text{F}_{visual}(image_{i}) \label{image_embedding_of_item}, v_{image}^{i} \in \mathbb{R}^{1 \times d_{v} }
\end{equation}
\begin{equation}
    v_{text}^{i} = \text{F}_{text}(text_{i}) \label{text_embedding_of_item}, v_{text}^{i} \in \mathbb{R}^{ 1 \times d_{t} }
\end{equation}

\subsection{Graph Modality}
In addition to the visual and textual information, we also model the relationships between items and behaviors as the third modality to elucidate the complicated item-behavior relationship. Typically, the graph modality of item-behavior relationships over multi-behaviors for the item $i$ is defined by Eq.~\ref{graph_embedding_of_item}, where $G_{i, B}$ is a sub-graph with the item $i$ and all observed behaviors $B$, and $\text{F}_{graph}(\cdot)$ is an encoder trained on the full item-behavior graph $\mathcal{G}_{I, B}$.
\begin{equation}
    v_{graph}^{i} = \text{F}_{graph}(\mathcal{G}_{i, B}), v_{graph}^{i} \in \mathbb{R}^{1 \times d_{g}}
    \label{graph_embedding_of_item}
\end{equation}

Moreover, the user behaviors such as \textit{view, add to cart} and \textit{purchase} can be represented in graph modality in the MBRS. In an LLM-based recommender system, the textual description of these behaviors could provide a textual modality for multi-behavior recommendations. However, the textual modality can only capture semantic information from behaviors while ignoring structural information. To tackle this shortage, we add graph modality $v_{graph}^{b}$ from the item-behavior graph we mentioned, which could be computed by encoding $\mathcal{G}_{I, b}$ by
\begin{equation}
v_{graph}^{b} = \text{F}_{graph}(\mathcal{G}_{I, b}), v_{graph}^{b} \in \mathbf{R}^{1 \times d_g}
\end{equation}

The graph modality incorporates the multi-behavior dependency into a graph architecture to capture the hierarchical item correlations in a customized manner. The encoder for the item-behavior graph can be any graph-based model that can yield the embeddings for both items and behaviors. In this paper, we use MBHT~\cite{yang2022multi} as the graph encoder in our TMF framework.

\subsection{LLM-based Recommender}
\textbf{LLM-based Recommender}. An LLM-based recommender for MBRS could be classified into three steps: (1) representation of user history, (2) prompt formatting, and (3) LLM inference for the next-item recommendation. 
The history of a given user $u$ consists of the past $n$ interacted items $H_{u} = {(i_1, b_1), (i_2, b_2), ..., (i_n, b_n)}$, where $i_n$ is the most-recently interacted items via behavior $b_n$. To represent $H_{u}$ in natural language, a common way is to convert items and the associated behaviors into their names and concatenate them into a long sentence $text_{u}$.
In the second part, the user history $text_{u}$ will be integrated into a prompt that an LLM can use to understand the context and instructions. 
Finally, an LLM processes the prompt and predicts the next item, i.e., the \text{n+1}-th item given $H_u$. 
To better align MBRS with LLMs, motivated by~\cite{liao2023llara}, we introduce special tokens after each item's name and behavior's name and initialize the embeddings of special tokens by projecting the item and behavior embeddings from graph-based MBRS model to the token space. 

\noindent \textbf{Instruction Tuning}. Instruction tuning has become a crucial technique to enhance the performance of Large Language Models (LLMs) on task-specific instructions. It significantly improves the ability of LLMs to follow human task-specific feedback~\cite{ouyang2022training}. This approach involves reorganizing data to define textual instructions and their corresponding responses clearly. By pairing task descriptions with their associated responses in a natural language format, a more comprehensive instructional context is created. Following this, the LLMs can be fine-tuned using the autoregressive objective.

\noindent \textbf{Parameter Efficient Fine-Tuning with LoRA.} Fine-tuning all the parameters of an LLM is a time-intensive process. To address this issue, Parameter-Efficient Fine-Tuning (PEFT)~\cite{DBLP:conf/iclr/HeZMBN22, DBLP:conf/emnlp/LesterAC21} focuses on optimizing a smaller subset of parameters, greatly reducing the computational demands while still maintaining strong performance. LoRA~\cite{DBLP:conf/iclr/HuSWALWWC22}, a common PEFT algorithm, keeps the LLM's weights fixed and decomposes the update process into trainable low-rank matrices. 

\section{Methodology}

In this section, we explain the modality fusion module to integrate visual, textual and graph data into 
an LLM. Then we introduce the prompt designs of the LLM and discuss the instruction tuning details.

\subsection{Modality Fusion Module} \label{adaptors}
To unleash the power of different modalities of items in MBRS, we employ pre-trained models to generate embeddings of items for each modality and introduce two mechanisms for modality fusion (Figure \ref{att}). 
In this three-level design, we first consider a self-attention mechanism on top of the user sequence with embeddings of all item modalities to learn better query embeddings for cross-modality attentions. For the second and third levels, we consider the output of the previous layer as a query and treat the image and the text as key and value (vice versa), respectively.

\subsubsection{All-Modality Self-attention (AMSA)}
The motivation behind the All-Modality Self-attention (AMSA) is that a proper query for cross-modality attention could be determined by both the modalities of an item and other items in the same user behavior sequence. 
We concatenate embeddings of different modalities of the item $i$ generated by their associated encoders to get a single representation of all modalities,
\begin{equation}
    v_M^i = \text{CONCAT}(v_{image}^{i}, v_{text}^{i}, v_{graph}^{i}), 
\end{equation}
where $v_M^i \in \mathbb{R}^{1 \times d}$, $d = \sum d_{m_j}$. Given a user behavior sequence, $H_u$, we can represent the sequence by stacking the item representation, $v^u_{M} \in \mathbb{R}^{|H_u| \times d}$.
Next, we use self-attention to address the dynamic importance of modalities during fusion (Eq. \ref{self_attention_adaptor}). In this design, the self-attention weight matrix ${\color{cyan}W_{att}} \in \mathbb{R}^{|H_u| \times |H_u|}$ can capture the dynamic importance of items in the user sequence. 

\begin{equation}
    v_{\text{amsa}}^{u} = \underbrace{softmax\left(\frac{v^u_{M} (v^u_{M})^T}{\sqrt{d_T}}\right)}_\text{\color{cyan}dynamic importance $W_{att}$}  v^u_{M}
    \label{self_attention_adaptor}
\end{equation}

\subsubsection{Cross-modality Attention (CMA)}
To further leverage the rich context in the image embedding and the text embedding, we build a cross-modality attention mechanism for modality fusion after AMSA. 
We consider the output of AMSA as the query to extract useful information from the image and the text modality. 
As illustrated in Figure~\ref{att}, we project the output of AMSA, $v^u_{amsa}$, into the same dimension of the image embedding and the text embedding, and apply the output embedding from AMSA as the query, and the image and text embeddings serve as the key and the value, respectively. 
For simplicity, we reuse $v^u_{amsa}$ for the query embedding after projection. We summarize the basic cross-modality attention adaptor in Eq. \ref{cross_attention_adaptor}, where $v_{image}^u \in \mathbb{R}^{|H_u| \times d_v}$ and $v_{text}^u \in \mathbb{R}^{|H_u| \times d_t}$, the image embeddings and the text embeddings of all items in the user sequence, respectively. 
\begin{equation}
    v_{\text{cma\_1}}^{u} = \underbrace{softmax\left(\frac{v_{\text{amsa}}^u  (v_{\text{image}}^u)^T}{\sqrt{d}}\right)}_\text{\color{cyan}dynamic importance $W_{text}$} v_{\text{text}}^{u}
    \label{cross_attention_adaptor}
\end{equation}
\begin{equation}
    v_{\text{cma\_o}}^{u} = \underbrace{softmax\left(\frac{v_{\text{cma\_1}}^u  (v_{\text{text}}^u)^T}{\sqrt{d}}\right)}_\text{\color{cyan}dynamic importance $W_{image}$} v_{\text{image}}^{u}
    \label{cross_attention_adaptor}
\end{equation}
We use each row of $v_{\text{cma\_o}}^{u} \in  \mathbb{R}^{|H_u| \times d_v}$ as the final item representation after cross-modality fusion. 


Note that both the all-modality self-attention and the cross-modality attention designs could be extended by their multi-head attention variants, boosting the attention performance \cite{vaswani2017attention}.

\subsection{Prompt Formatting}
In an LLM-based recommender, the user-item interaction sequence determines the context of user preference in the prompt.
Integrating different modalities of items and behaviors into a prompt for LLM-based recommenders requires a unique prompt design. Given an item and its associated behavior $(i, b) \in H_{u}$ for a user $u$, we represent them by their names, $(name_{i}, name_{b})$.
Instead of direct insertion of $(name_{i}, name_{b})$ into a prompt, we design a hybrid prompt and introduce extra modality tokens for items and behaviors after their names, respectively. 

Typically, we define the item $i$'s modality token as `$[MT_{i}]$', and its associated behavior $b$'s modality token as `$[MT_{b}]$'. These extra tokens are not seen during the pre-training steps of LLMs.
To warm up the embeddings of the behavior token  `$[MT_{b}]$' and the item ID token `$[MT_{i}]$', we develop two adaptor modules that project behavior embeddings and the output of the item modality fusion module (section \ref{adaptors}) to the LLM token embedding space, respectively.
We use a two-layer MLP to complete the projection.
Figure \ref{tmf-framework} shows the high-level design of our hybrid prompt. The language tokens are from the pre-trained LLM, while the item ID token `$[MT_{i}]$' and the behavior token `$[MT_{b}]$' are also included with embedding warm-up from adaptors. 

Instruction of the recommendation task is also critical for a successful LLM inference. We follow the prompt design in LLaRA \cite{liao2023llara} and include the candidate set into the prompt beside the task definition and the user-item interaction sequence.  The recommendation task is to predict the name of the next \textit{item} under a certain \textit{target user behavior}, given the candidate item set.




\subsection{Instruction Tuning with LoRA}
To properly align the knowledge in the pre-trained LLMs and the modality information from the user-item interactions, we conduct the instruction tuning with LoRA~\cite{DBLP:conf/iclr/HuSWALWWC22} for a pre-trained LLM over the hybrid prompt. However, because the LLMs are predominantly pre-trained on textual data, the tuning task complexity could increase if we include more modalities in the prompts. 
We consider the curriculum tuning strategy~\cite{liao2023llara} to determine the order of various tuning tasks. 

\begin{figure*}
    \centering
    \includegraphics[width=0.95\textwidth]{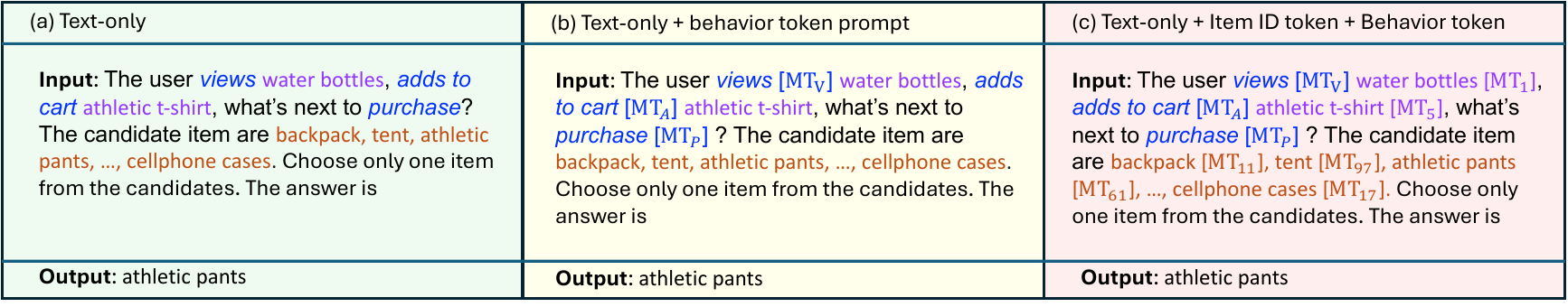}
    \caption{Task complexity and prompt examples: (a) the text-only task only uses the text prompt with item names and behaviors for user context. (b) and (c) are more challenging tasks to gradually introduce more modalities and tokens into the prompts.}
    \label{prompt-template-tasks}
\end{figure*}

\noindent\textbf{Task Complexity}. 
The task complexity is related to the heterogeneity of modalities in the prompts. 
We design learning tasks with three levels of complexity. The easy task has text-only prompts without introducing the behavior tokens and the item ID tokens (Figure \ref{prompt-template-tasks}(a)). The medium task considers both text-only prompts and the behavior tokens (Figure \ref{prompt-template-tasks}(b)). Finally, the hard task includes the item ID tokens on top of the medium task (Figure \ref{prompt-template-tasks}(c)). 

Let $\Phi_{LLM}$ denote the LoRA parameters for LLM fine-tuning, and $\Psi_{f}, \Psi_{i}, \Psi_{b}$ be the learnable parameters of the modality fusion module, the adaptor for item ID tokens and the adaptor for behavior tokens, respectively. We list the loss function for each task for optimization in Eq. \ref{easy}-\ref{hard}, where $(H_u, y_u)$ is the input behavior sequence and the target interaction pair. We only keep the trainable parameters for simplicity. 
\begin{align}
    \mathcal{L}_{easy} (H_u, y_u) &= \sum^{|y_u|}_{t=1} \log P_{[\Phi_{LLM}]}(y_{u,t} | H_u, y_{u, <t}) \label{easy}\\
    \mathcal{L}_{medium} ((H_u, y_u) &= \sum^{|y_u|}_{t=1} \log P_{[\Phi_{LLM}+\Psi_{b}]}(y_{u,t} | H_u, y_{u, <t}) \label{medium}\\
    \mathcal{L}_{hard} ((H_u, y_u) &=  \sum^{|y_u|}_{t=1} \log P_{[\Phi_{LLM}+\Psi_{b}+\Psi_{f}+\Psi_{i}]}(y_{u,t} | H_u, y_{u, <t}) \label{hard}
\end{align}

\noindent\textbf{Scheduler Formulation and Training}. 
To transfer the learning from the easy task to more challenging tasks, we introduce two thresholds on the number of training iterations, $T_1 < T$, to control the transition from easy to medium, where $T$ is the total number of training iterations. Typically, we define $p_{m}(\tau)$ as the probability of picking the medium task to learn from only the easy and the medium task in Eq. \ref{p_medium}, where $\tau$ represents the iterations. As $\tau$ increases, the learning system should give a higher probability of assigning the medium task, and $1 - p_{m}(\tau)$ is the probability of assigning the easy task. 
\begin{equation}
    p_{m}(\tau | \{E, M\}) = \frac{\tau}{T}, 0 \leq \tau \leq T_1 \label{p_medium}
\end{equation}
Once $\tau \geq T_1$, we start introducing the hard task. We define the probability of choosing the hard task from on the medium and the hard task $p_{h}(\tau | \{M, H\})$ definition in Eq. \ref{p_medium_total}, and $1-p_{h}(\tau | \{M, H\})$ is the probability of assigning the medium task in this case. 
\begin{equation}
     p_{h}(\tau | \{M, H\}) = \frac{\tau - T_1}{T-T_1}, T_1 \leq \tau \leq T \label{p_medium_total}
\end{equation}

The final loss for optimization could be formulated under different learning stages Eq. \ref{stage_2}, where $\mathbb{I}_{(a, b)}$ is an indicator of whether a random variable from the uniform distribution has the probability value between $a$ and $b$. 
\begin{equation}
   \mathcal{L}_{\Phi, \Psi}= \left\{
\begin{array}{ll}
      \sum_{\tau} \mathbb{I}_{(p_{m}, 1]}\mathcal{L}_{easy} + \mathbb{I}_{[0, p_{m}]} \mathcal{L}_{medium} & 0\leq \tau \leq T_1 \\
      \sum_{\tau} \mathbb{I}_{(p_{h}, 1]}\mathcal{L}_{medium} + \mathbb{I}_{[0, p_{h}]} \mathcal{L}_{hard} & T_1< \tau \leq T  
\end{array} 
\right.  
\label{stage_2}
\end{equation}

\noindent \textbf{LLM-based Generation Process. }
As LLMs parse the hybrid prompt with user history data, the inference process predicts the subsequent tokens as the name of the predicted item. 
The overall generation process of the TMF framework for multi-behavior recommendation is explained in Figure \ref{tmf-framework}. The behavior tokens and the item ID tokens won't be part of the predicted tokens during generation.

\section{Experiments}

\subsection{Datasets and Setup}
\noindent \textbf{Datasets.} We consider {\color{black}real-world customer behaviors from Walmart e-commerce platform} and sample data from three categories: (1) Electronics, (2) Pets, and (3) Sports. We consider \textit{view}, \textit{add to cart}, and \textit{purchase} to represent customer behaviors on items. We drop the customer behavior sequence without a \textit{purchase} behavior and convert data from each category to a dataset. Table \ref{dataset} summarizes the statistics of our datasets. We also report the `\#Item/\#User Ratio' to indicate the user behavior complexity for each dataset: the higher ratio means users interact with more diverse set of items and show more complicated shopping intents in this dataset. 

\begin{table}[]
    \centering
    \caption{Data Description}
    \label{tab:data_stats}
    \begin{tabular}{lccc}
    \hline 
         & Electronics & Pets & Sports \\
         \hline 
         \#Items & 48270 & 25007 & 28558\\ 
         \#Users & 5091 & 17228 & 4488 \\ 
         Average Length & 23 & 20 & 24\\ 
         Behavior Types & 3  & 3 & 3\\ 
         \#Item/\#User Ratio & 9.481 & 1.452 & 6.363 \\
         \hline
    \end{tabular}
    \label{dataset}
\end{table}

\noindent \textbf{Implementation Details.} We select Llama2-7B \cite{Touvron2023Llama2O} as the backbone model in the LLM-based Recommender. To be compatible with the curriculum learning design, the instruction format for training and testing is randomly sampled from several prompts. 
The pre-trained encoders for item images and textual information are all from  CLIP~\cite{radford2021learning}. 
The MHBT model~\cite{yang2022multi} generates the item embeddings and behavior embeddings learned from the item-behavior graph.
For all LLM-related methods, each experiment is trained for a maximum of 3 epochs, with a batch size of 64 on one A100 GPU. We follow the setting in \cite{liao2023llara} for learning rate warm-up and grid search. We use Adam Optimizer to optimize the adapter parameters and the LoRA parameters of LLM layers. 

\noindent \textbf{Evaluation Metrics.} For each user and the associated behavior sequence, we randomly select
10 non-interacted items to construct the candidate set, and the ground truth is included in the candidate set. TMF and other baseline models are evaluated on identifying the ground truth item out of the candidate set. Their performance is evaluated using the \textit{HitRate@1} metric.
In addition to the HitRate@1 metric, we report another metric called \textit{valid ratio} \cite{liao2023llara} to evaluate the capability of instruction following during generation. It computes the proportion of valid responses (i.e., items in the candidate set) across all behavior sequences. Please note that the valid ratio of non-LLM models will always be 1.0 because they are embedding-based methods and no instructions are involved when recommending items.

\begin{table*}[]
    \centering
    \caption{The Results of TMF compared with the sequential recommender models, and LLMs-based methods.}
    \label{tab:result}
\begin{tabular}{lllllll}
\hline
   & \multicolumn{2}{c}{\textbf{Electronics}} & \multicolumn{2}{c}{\textbf{Pets}} & \multicolumn{2}{c}{\textbf{Sports}} \\
   & HitRate@1             & ValidRatio             & HitRate@1              & ValidRatio             & HitRate@1              & ValidRatio             \\
   \hline
   GRU4Rec & 0.117             & 1.000              & 0.126              & 1.000              & 0.119             & 1.000             \\
SASRec & 0.123             & 1.000              & 0.131              & 1.000              & 0.123             & 1.000             \\
Bert4Rec & 0.186             & 1.000              & 0.194              & 1.000              & 0.179             & 1.000             \\
MBGCN & 0.473             & 1.000              & 0.571              & 1.000              & 0.559             & 1.000             \\
MBRec & 0.508            & 1.000              & 0.592              & 1.000              & 0.575             & 1.000             \\
MBHT & 0.540           & 1.000              & 0.621              & 1.000              & \underline{0.601}             & 1.000             \\
LLaRA &0.602 & 0.660 & 0.837 & 0.945 &0.584 & 0.742 \\
LLama-2 & \underline{0.617}              & 0.628              & \underline{0.859}           &  0.997               & 0.490             & 0.788             \\
TMF & \textbf{0.853} (+38.250\%)             & 0.998              &\textbf{0.885 } (+3.027\%)             & 0.968              & \textbf{0.860} (+43.095\%)             & 0.986   \\
\hline
\end{tabular}
\end{table*}

\begin{table*}[]
    \centering
    \caption{The Results of TMF under different levels of Modality Fusion (Ablation study).}
    \label{tab:ablation}
\begin{tabular}{lllllll}
\hline
   & \multicolumn{2}{c}{\textbf{Electronics}} & \multicolumn{2}{c}{\textbf{Pets}} & \multicolumn{2}{c}{\textbf{Sports}} \\
   & HitRate@1             & ValidRatio             & HitRate@1              & ValidRatio             & HitRate@1              & ValidRatio             \\
   \hline
TMF (Llama-2 7B) & 0.617              & 0.628              & 0.859           &  0.997               & 0.490             & 0.788             \\
+ Behavior Tokens & 0.648           & 0.718              & 0.860          &  0.998               & 0.557             & 0.745             \\
+ Item ID Tokens (AMSA) & 0.823              & 0.983              & 0.865           &  0.997               & 0.817            & 0.974             \\
+ CMA layers & \textbf{0.853} (+3.645\%)         & 0.998              &\textbf{0.885 }(+2.312\%)             & 0.968              & \textbf{0.860} (+5.263\%)             & 0.986   \\
\hline
\end{tabular}
\end{table*}

\begin{table*}[]
    \centering
    \caption{Human Rating Scoare and Inter-rater Reliability (Fleiss’ Kappa). All Kappa statistics have p-value $\leq 0.01$.}
    \label{tab:rating}
\begin{tabular}{lllllll}
\hline
   & \multicolumn{2}{c}{\textbf{Electronics}} & \multicolumn{2}{c}{\textbf{Pets}} & \multicolumn{2}{c}{\textbf{Sports}} \\
   & Avg. Rating             & Kappa             & Avg. Rating             & Kappa             & Avg. Rating             & Kappa           \\
   \hline
MBHT & 3.845              & 0.284              & 3.667           &  0.198               & 3.313            & 0.249             \\
TMF & \textbf{4.603} (+19.713\%)         & 0.0637              &\textbf{4.403}(+20.071\%)             & 0.149              & \textbf{4.157 } (+25.475\%)             & 0.185   \\
\hline
\end{tabular}
\end{table*}

\subsection{Comparison Experiments}
\noindent \textbf{Baselines. }  We mainly consider three categories of baseline models:
\begin{list}{$\square$}{\leftmargin=1em \itemindent=0em}
  \item \textit{Traditional Sequential Recommender System}: 
  
  \textbf{GRU4Rec}~\cite{hidasi2015session}, \textbf{SASRec}~\cite{kang2018self} and \textbf{Bert4Rec}~\cite{sun2019bert4rec} apply Recurrent Neural Network (RNN), attention mechanism and Bert~\cite{devlin2018bert} to model user behavioral sequences. 

  \item \textit{Graph-based Multi-behavior Recommender Systems}: 
  
  \textbf{MBGCN}~\cite{jin2020multi}, \textbf{MBRec}~\cite{xia2022multi} and \textbf{MBHT}~\cite{yang2022multi} utilize is a multi-behavior graph neural network and hypergraph enhanced transformer to learn the cross-type behavior interaction patterns.
  \item \textit{LLM-based recommender system}: 
  
  \textbf{LLaRA}~\cite{liao2023llara} uses a hybrid prompting method with textual item features that integrates ID-based item embeddings learned by SASRec.

  \textbf{Llama-2} \cite{Touvron2023Llama2O} is an open-source LLM released by Meta. In our experiments, the proposed TMF with Llama2 7B backbone can be reduced to Llama-2 if no modalities from items and behaviors are considered. We use the reduced Llama-2 with the same prompts with TMF as a baseline and fine-tune it on our datasets.

\end{list}

\noindent \textbf{Experiment Results. } We summarized the experiment results in Table \ref{tab:result}. We highlight the best scores in bold and the second-best scores with underline. From Table~\ref{tab:result}, we have the following observations:
\begin{itemize}
    \item We can see that TMF outperforms all other baselines on all datasets. For both Electronics and Sports datasets, our model achieves more than 38\% on HitRate@1. Note that Electronics and Sports datasets have very high \#Item/\#User ratio which indicates complicated user-item interactions. These numbers demonstrate the effectiveness of TMF in modality fusion on multi-behavior recommendation tasks under different levels of user-item interaction complexity. 
    \item  TMF outperforms all LLM-based recommender baselines after introducing the modality fusion. For the Pets dataset, TMF still outperforms Llama-2 baseline which doesn't contain any modality information except the basic user context. This is because the Pets dataset has the lowest \#Item/\#User ratio and users behaviors are less complicated compared with other datasets, which could be learned by Llama-2 over our fine-tuning. However, our TMF with modality fusion could further improve the performance with a similar valid ratio for generation quality.
\end{itemize}


\begin{figure*}[h]
    \centering
    \includegraphics[width=\textwidth]{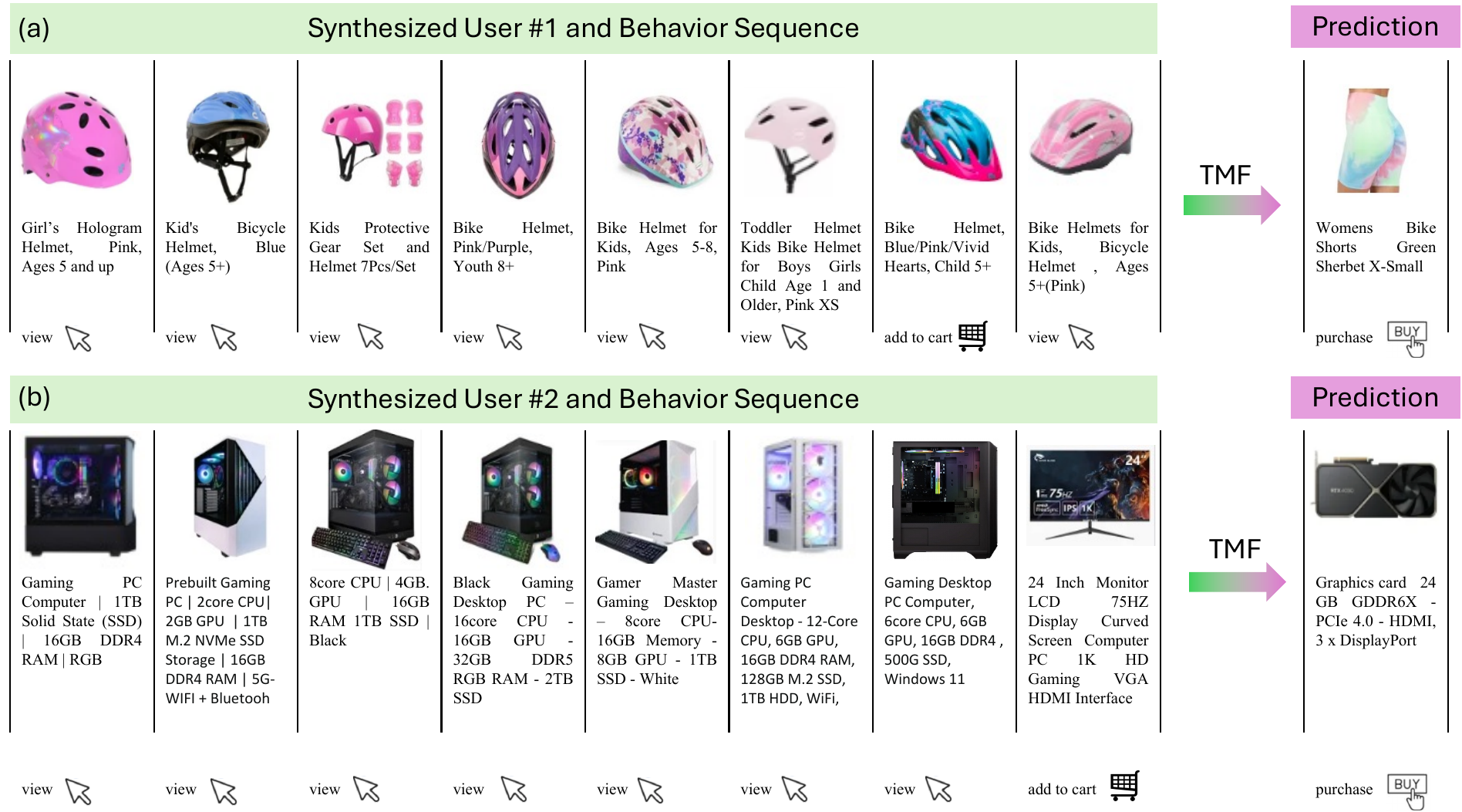}
    \caption{Case Study on Sports and Electronics datasets to demonstrate the TMF's capacity on context understanding (shopping topics, user age and gender requirement, design, etc.) and reasoning for next purchase prediction.}
    \label{case_study}
\end{figure*}

\subsection{Ablation Study}
To better understand the effectiveness of different modules in our model, we conduct ablation studies starting from the basic TMF (Llama-2 7B) model, which is only the Llama-2 backbone of the TMF. By adding the behavior tokens, Item ID tokens (AMSA) and CMA layers one by one into the basic TMF, we have three more ablation models and the last one is the complete TMF. The results of the ablation studies are presented in Table~\ref{tab:ablation}. From this table, we find:
\begin{itemize}
    \item Across all three datasets, the full TMF model consistently outperforms, while the basic TMF (Llama-2 7B) shows the weakest performance. Each additional module contributes to improved model performance, highlighting the effectiveness of the Behavior tokens, item ID tokens (AMSA), and the CMA layer. Compared to Llama-2, TMF offers a richer representation of items and behaviors, significantly enhancing the LLM's ability to make accurate recommendations to users.
    \item Among all three modules, the item ID tokens (AMSA) provide the most substantial improvement in model performance. This suggests that incorporating multiple modalities of item information into the model allows the LLM-based recommender to better understand the items. Additionally, our proposed AMSA module effectively integrates image, text, and graph data and projects them into the LLM, enabling the model to leverage this rich information to enhance its recommendation capabilities.
\end{itemize}

\subsection{Human Evaluation}
Our TMF framework has been deployed in a real-world recommendation production to generate candidate sets to ranking models. To understand the improvement of the TMF in real-world production, we leverage human raters to evaluate recommendations from TMF and the best non-LLM-based sequential recommender model MBHT under the multi-behavior setting. We sample 100 user behavior sequences from each dataset and ask 15 human raters to score the recommendation from TMF and MBHT, respectively. Raters must consider the given behavior sequences as the context to evaluate their likelihood of purchasing the recommendation. The score ranges from 1 to 5, 1 for totally irrelevant purchases and 5 for tightly relevant purchases. Each rater needs to score all 100 cases for both TMF and MBHT without knowing the source of recommendations (i.e., from MBHT or TMF) and the ground truth labels. 

We summarize the rating score and report the inter-rater reliability scores (Fleiss’ Kappa) in Table \ref{tab:rating}. We can see that TMF achieves better rating scores in all three datasets than MBHT. Because human raters don't know the model sources of the recommendations and the ground truth, the recommendations from TMF under the multi-behavior setting have higher quality and better acceptance rates by humans. This demonstrates that TMF with LLM backbone and the multi-modality fusion could understand richer sequence contexts for more reliable recommendations.   

\begin{figure}
    \centering
    \includegraphics[width=0.49\textwidth]{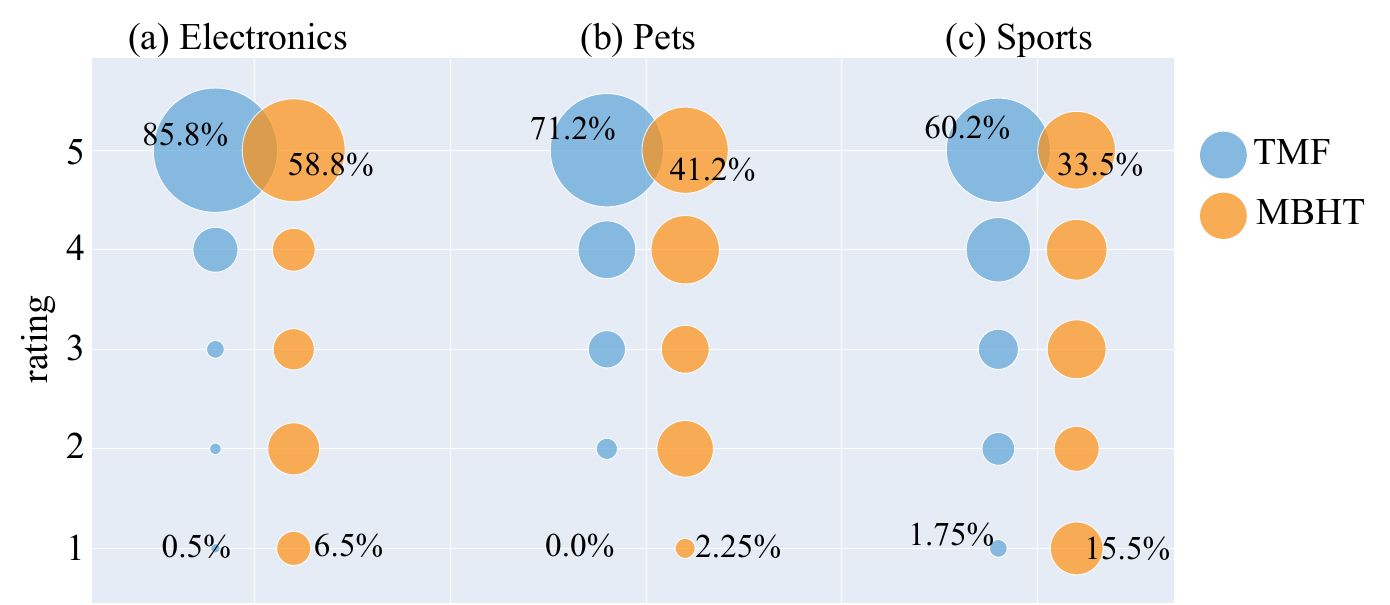}
    \caption{Distribution of Human Rating Scores in a bubble chart.}
    \label{all_rating_dist}
\end{figure}

To understand the distribution of rating scores for each model, we plot a bubble chart in Figure \ref{all_rating_dist}. In this bubble chart, the y-axis is the rating scale, 1 to 5, and the size of bubbles represents the percentage of a certain rating score in the total rating results for a given (model, dataset) pair. Most of the rating scores for recommendations from TMF (blue bubbles) fall into the best rating, while MBHT (orange bubbles) shows a considerable amount of lower ratings. This demonstrates the stable performance of TMF in generating high-quality recommendations. This matches the pattern shown in Table \ref{tab:rating}. 
Note that TMF also has less neutral rating scores than MBHT, which unveils TMF's strong capacity in context understanding and recommendation intent clarity.

\subsection{Case Studies on Synthetic Data}
We conducted the case study on the Sports and Electronics datasets to illustrate the recommendation generated by the TMF.
We synthesize two user behavior sequences under Electronics and Sports, respectively, and use TMF to predict the next purchase. The prediction targets are also curated carefully by human to reflect the underlying relevance within each modality between the prediction and the user behavior sequence.
Typically, we follow these principles to simulate the real shopping behaviors:
\begin{list}{$\square$}{\leftmargin=1em \itemindent=0em}
\item \textit{\textbf{Modal-specific Preference}}: Item images and textual descriptions reflect (explicitly or implicitly) the user's preference on the visual design (i.e., color, pattern, design, style) and textual detials such as brand, size, age and gender, respectively. 
\item \textit{\textbf{Consistency in User Profile}}: User profiles, such as demographics, are usually consistent over multiple behaviors. 
\item \textit{\textbf{Shopping Intent Awareness}}: User behavior sequences and targets should be relevant and under the same shopping topic. 
\end{list}
We present the user behavior sequences and predictions in Figure~\ref{case_study}.

We find TMF has the ability to generate different item types from the user's history. In Figure~\ref{case_study} (a), User\#1 only views and adds to cart colorful bike helmets for children, while TMF recommends a colorful X-small women bike shorts. This example shows that TMF can understand the user profile (young girl) and user intent (colorful bicycle equipment). Note that our modality fusion design helps TMF capture the requirement on the pinkish color of the recommendation, which is only indicated by the image of the bike shorts rather than its description. 
Besides, the recommendation is not limited to the user's interacted item type. Instead, TMF finds bike shorts, which are complementary to bike helmets. It indicates the reasoning capability of TMF, which may come from the backbone LLM~\cite{brown2020language}.

We also find that TMF can predict the next item with great shopping intent awareness by using its extensive knowledge. In Figure~\ref{case_study} (b), the User\#2 browses multiple gaming PC computers and one PC monitor while TMF recommends an advanced graphics card. Though gaming PC computers are rarely purchased together with game PC computers because gaming PC computers always have a graphics card inside, the advanced graphics card can upgrade the user's current computer. Therefore, the recommended graphics card can be viewed as a substitute for gaming PC computers. We assume this knowledge is from the LLM which is pre-trained on massive datasets.


\section{Conclusion}
In this paper, we introduce the Triple Modality Fusion (TMF) framework to enhance multi-behavior recommendation systems (MBRS) by integrating visual, textual, and graph modalities with the power of large language models (LLMs). The inclusion of diverse data types allows the model to capture a multifaceted understanding of user behaviors and item characteristics, offering a more accurate and contextually relevant recommendation results. We also propose the modality fusion module based on self-attention and cross-attention mechanisms. Our extensive experiments demonstrate the TMF framework's superior performance in improving recommendation accuracy. Additionally, ablation studies affirm the critical role of all modalities and the effectiveness of the cross-attention mechanism used in modality fusion. 


\bibliographystyle{ACM-Reference-Format}
\balance
\bibliography{sample-base}

\end{document}